\documentclass{webofc}
\usepackage[varg]{txfonts}   % Web of Conferences font
\begin{document}
\title{The impact of isospin dependence of pairing on fission
barriers in the fission cycling regions}

\author{\firstname{A.\ V.} \lastname{\ Afanasjev}\inst{1}\fnsep\thanks{\email{aa242@msstate.edu}} \and
        \firstname{A.} \lastname{\ Taninah}\inst{1}    % etc.
}

\institute{Department of Physics and Astronomy, Mississippi
State University, MS 39762          }

\abstract{
   A systematic analysis of the ground state and fission properties of actinides and 
superheavy nuclei  important for the $r$ process modeling has been performed 
within the framework of covariant density functional theory for the first time in Ref.\ 
\cite{TAA.20}. A brief review of the results related to the heights of primary fission barriers and systematic 
uncertainties in their prediction is presented. In addition, new results on the potential 
impact of the isospin dependence of pairing on fission barriers in fission cycling
regions is provided for  the first time. 
}
\maketitle

%%%%%%%%%%%%%%%%%%
\section{Introduction}
\label{intro}
%%%%%%%%%%%%%%%%%%

   The $r$ process is responsible for the synthesis of approximately half of the nuclei in nature 
beyond Fe \cite{P.08} and it is the only process which leads to the creation of nuclei heavier than Bi 
\cite{LMHCF.18} . 
%The modeling of the $r$ process in neutron-rich environments (such as neutron star mergers) 
%depends sensitively on nuclear masses, $\alpha$- and $\beta$-decay half-lives, neutron capture, 
%and fission properties of the nuclei, the majority of which will never be measured in laboratory 
%conditions [2,4,5,10].  
Fission becomes important in the $r$ process simulations for the  neutron-to-seed ratios which 
are large enough to produce fissioning nuclei \cite{MMZKLPPRST.07,G.15}. The
$r$ process can reach the region beyond neutron shell closure at $N=184$ for these ratios
exceeding 100: the fission plays a dominant role in this region.  Fission leads to the termination 
of the hot $r$ process by means of {\it fission cycling} which returns matter to lighter nuclei 
\cite{MMZKLPPRST.07,G.15}. It also determines the strength of fission cycling, the ratio of the actinides to light and medium 
mass $r$ process nuclei, and thus the shape of the final element abundance pattern. In addition,
it defines the possibility of the formation of neutron-rich superheavy nuclei in the $r$ process 
\cite{PLMPRT.12}.

   The outcome of the r-process modeling sensitively depends on the quality of employed theoretical 
frameworks and associated theoretical uncertainties and their propagation on going to neutron-rich 
nuclei in the situation when experimental data are not known. So far only non-relativistic frameworks 
have been used in such modeling and in the analysis of fission cycling (see review in the introduction of Ref.\
\cite{TAA.20}). The first attempt to produce nuclear input required  for the $r$ process modeling within 
the relativistic framework (covariant density functional theory (CDFT) \cite{VALR.05}) has
been carried out in Ref.\ \cite{TAA.20}. The goal of this contribution is to briefly review the results
of this study with major focus on fission properties and to analyze potential  impact of isovector
dependence of pairing on the fission barriers of very neutron-rich nuclei.

%%%%%%%%%%%%%%%%%%%%%%%%%%%%%%%%
\section{Fission barriers and related systematic uncertainties}
\label{sec-2}
%%%%%%%%%%%%%%%%%%%%%%%%%%%%%%%%

%%%%%%%%%%%%%%%%%%%%%%%%%%%%%%%%%%%%%%%%%%%
\begin{figure*}[htb]
\centering
\includegraphics[angle=-90,width=14cm,clip]{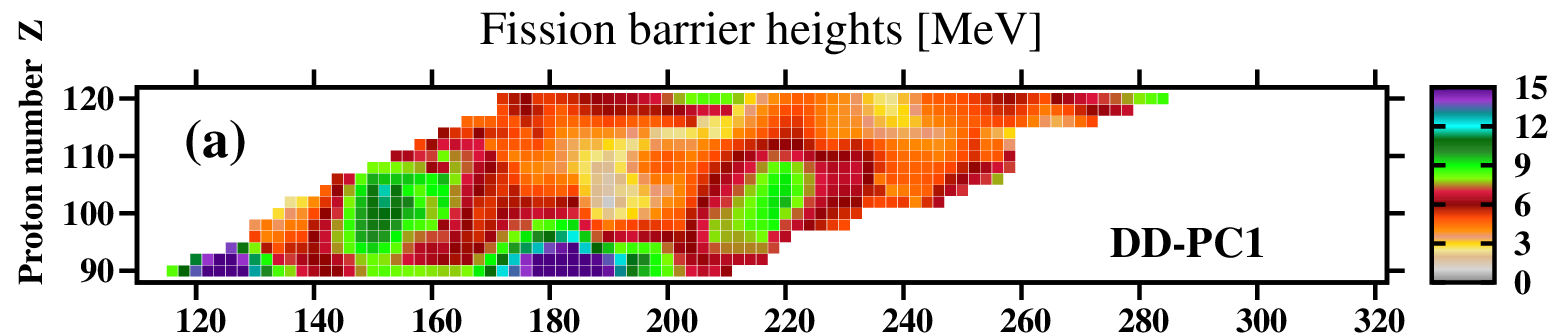}
\includegraphics[angle=-90,width=14cm,clip]{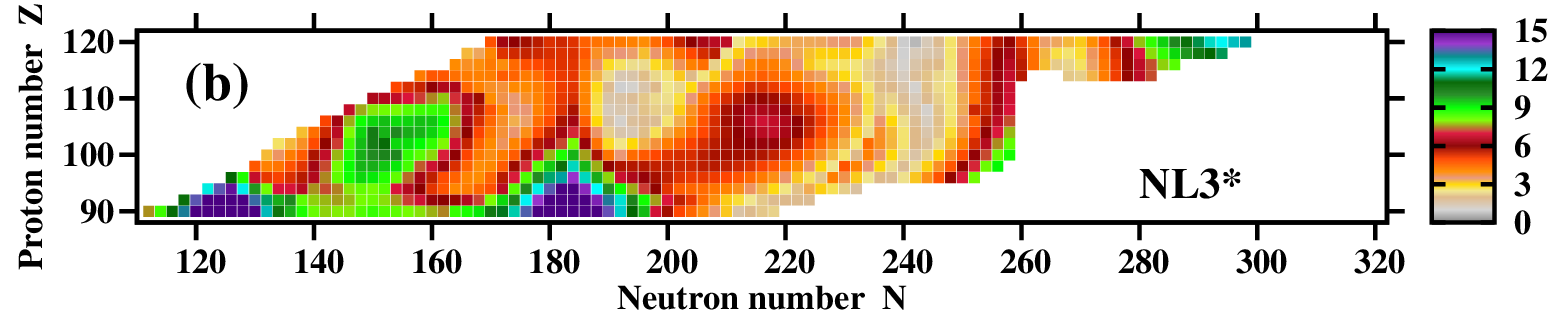}
\caption{
The heights $E^B$ (in MeV) of PFBs  obtained in axial RHB calculations as 
a function of proton and neutron numbers for nuclei located between two-proton and 
two-neutron drip lines. Based on Fig. 15 of Ref.\ \cite{TAA.20}. 
\label{fig-EB}       
}
\end{figure*}
%%%%%%%%%%%%%%%%%%%%%%%%%%%%%%%%%%%%%%%%%%

  The distributions of primary fission barrier (PFB)  heights in the $(Z,N)$ plane obtained 
in the axial relativistic Hartree-Bogoliubov (RHB) calculations are shown in Fig.\ 
\ref{fig-EB}. The calculations have been performed with four state-of-the-art
covariant energy density functionals (CEDFs) DD-PC1, DD-ME2, NL3* and PC-PK1 
in order to evaluate systematic theoretical uncertainties in the predictions of 
fission barriers (see Ref.\ \cite{TAA.20} for details). There is a large similarity 
of the results obtained with DD-PC1 and DD-ME2 on the one hand and
NL3* and PC-PK1 on the other: thus only those obtained with DD-PC1
and NL3* are shown. In general, the topologies of the fission barrier maps shown
in Fig.\ \ref{fig-EB} are similar.  However, the fission barriers obtained with
DD-PC1 are on average higher by approximately 2 MeV as compared with
those calculated with NL3*. This is a consequence of different nuclear matter
properties of these two functionals (see discussion in Ref.\ \cite{TAA.20}). 
These two functionals also differ with respect of the predictions of the formation 
of superheavy elements with $N>240$ in the $r$ process. This is because
the band of the nuclei around $N\approx 240$ has extremely low fission barriers 
with heights of around 2 MeV in the NL3* CEDF (see Fig.\ \ref{fig-EB}(b)). The 
nuclear flow during most of the neutron irradiation step of  the $r$ process follows 
the neutron drip line and this  flow will most likely be terminated at $N\approx 240$ 
nuclei because of these low fission barriers in the calculations with NL3*.

\begin{figure*}[h]
\includegraphics[angle=-90,width=14cm,clip]{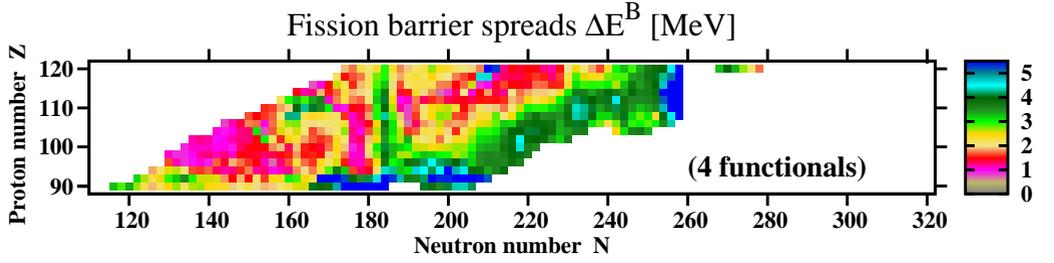}
\caption{
The spreads $\Delta E^B$ of the heights of PFBs as a function of proton and
%The spreads  $\Delta E^B$ of the heights of PFBs as a function of proton and 
neutron numbers. $\Delta E^B (Z,N) = |E^B_{max}(Z,N) - E^B_{min}(Z,N) |$, where, for given
%neutron numbers.  $\Delta E^B (Z,N) = |E^B_{max}(Z,N) - E^B_{min}(Z,N)|$, where, for given 
$Z$ and $N$ values, $E^B_{max}(Z, N)$ and $E^B_{min}(Z,N)$ are the largest and smallest heights 
of PFBs obtained with the employed set of four functionals. Based on Fig. 15 
of Ref.\ \cite{TAA.20}.
\label{fig-delta-EB}     
}
\end{figure*}

  Theoretical systematic uncertainties in the heights of PFBs given by the spreads 
$\Delta E^B$ are shown as a function of proton and neutron numbers in Fig.\   
%$\Delta E^B$  are shown as a function of proton and neutron numbers in Fig.\ 
\ref{fig-delta-EB}.   They are relatively modest in some regions but are enhanced near 
the $N=184$ and $N=258$ shell closures, for the $Z\approx 90$ 
nuclei with $N=166-184$ and in the wide band of nuclei parallel to the 
two-neutron drip line. The analysis of these spreads allows us to identify two 
major  sources of theoretical uncertainties in the predictions of the heights of PFBs
(see Ref.\ \cite{TAA.20} for details).  These are underlying single-particle structure (especially the one 
in the vicinity of shell closures) and nuclear matter properties of employed CEDFs. The 
former mostly affects the predictions for the ground states (and thus for the heights of PFBs) 
in the first two regions and  the latter the predictions for PFBs of the nuclei  located in the 
vicinity of the neutron drip line.

%%%%%%%%%%%%%%%%%%%%%%%%%%%%%%%%%%%%%%%%
%%%%%%%%%%%%%%%%%%%%%%%%%%%%%%%%%%%%%%%

%%%%%%%%%%%%%%%%%%%%%%%%%%%%%%%%%%%%%%
\section{The impact of isospin dependence of pairing on fission barriers}
\label{sec-3}
%%%%%%%%%%%%%%%%%%%%%%%%%%%%%%%%%%%%%%

%%%%%%%%%%%%%%%%%%%%%%%%%%%%%%%%%%%%%%%%%%%%%
\begin{table}
\centering
\caption{Different versions (v1, v2 and v3)  of separable pairing interaction as defined 
by the particle number dependencies of their scaling factors $f_i$ ($i=\pi$ or $\nu$). 
The constants $C_i$ and $\alpha_i$ are taken from Table 4 of Ref.\ \cite{TA.21}.}
\label{tab-pair-def}   
\begin{tabular}{|c|c|c|c|}
\hline
subsystem   &     v1               &    v2                                                                                  &    v3                                                           \\  \hline
   proton       & $f_{\pi}=1.0$  &   $f_\pi = C_{\pi}*(N+Z)^{\alpha_\pi}$                              &   $f_\pi = C_{\pi}*e^{\alpha_\pi|N-Z|}$           \\  
   neutron     & $f_{\nu}=1.0$ &   $f_\nu = C_{\nu} e^{\alpha_\nu\frac{|N-Z|}{N+Z}}$      &   $f_\nu= C_\nu*|N-Z|^{\alpha_\nu}$               \\ 
                    &                       &  $C_\pi=1.877$, $\alpha_\pi=-0.1072$                           &   $C_\pi=1.178$, $\alpha_\pi=-0.0026$     \\ 
                    &                       &   $C_\nu=1.208$, $\alpha_\nu=-0.674$                          &   $C_\nu=1.264$, $\alpha_\nu=-0.0495$  \\ \hline
\end{tabular}
\end{table}
%%%%%%%%%%%%%%%%%%%%%%%%%%%%%%%%%%%%%%%%%%%%%%

 The systematic calculations of Ref.\ \cite{TAA.20},  the results of which are presented in Figs.\ \ref{fig-EB}
and \ref{fig-delta-EB}, have been performed with separable pairing of Ref.\ \cite{TMR.09} 
the scaling factors $f_{i}$ of which  for the nuclei  with $Z>88$ are set to $f_{\pi}=f_{\nu}=1.0$ (see Ref.\ 
\cite{AARR.14} for definition of pairing strength).
This pairing is  labeled as "v1" below, see Table \ref{tab-pair-def}. However, in general more complicated 
particle number dependencies of  scaling factors $f_{i}$ are allowed  (see Refs.\ \cite{AARR.14,TA.21}). 
Indeed,  recent systematic analysis of Ref.\  \cite{TA.21} clearly reveals isospin dependence of scaling 
factors $f_{\nu}$ of neutron pairing.  However, the situation is less certain in the proton subsystem since 
similar accuracy of the description of pairing indicators can be achieved both with isospin-dependent 
and mass-dependent scaling factors $f_{\pi}$.

 The fission barrier heights sensitively depends on the strength of pairing interaction (see Ref.\ 
\cite{KALR.10}). Thus, it is important to understand by how much and in which direction the fission 
barriers can be affected by these particle number dependencies of pairing interaction. For that
the primary fission barriers of very neutron-rich $^{316}$Fm and $^{324}$Rf nuclei have been 
calculated with three versions (v1, v2 and v3) of scaling factors for separable pairing (see Table  
\ref{tab-pair-def}). These nuclei are located at the neutron-rich side of expected fission cycling 
region (see Fig. 1 in Ref.\ \cite{TAA.20}). The versions v2 and v3 are the combinations of proton and 
neutron scaling factors favored by the analysis of Sec. IV of Ref.\ \cite{TA.21}.  One can see in Table 
\ref{tab-FB}  that in all cases  the inclusion of isospin dependence of pairing increases the 
heights of primary fission barriers but the magnitude of the increase depends both on the CEDF 
and nucleus. The latter feature is mostly due to the differences between the functionals in underlying 
single-particle structure. The magnitude of the increase varies from 
$E^B_{\rm v2} -E^B_{\rm v1} = 0.25$ MeV  to $E^B_{\rm v3} -E^B_{\rm v1} = 1.57$ MeV (see 
DD-ME2 results in $^{324}$Rf). This large difference in $^{324}$Rf is caused by substantially 
weakened proton pairing in the calculations with the v3 version of separable pairing which according 
to Ref.\ \cite{KALR.10} leads  to a significant increase of fission barrier height. However,  in most of 
the cases the difference between $E^B_{\rm v2} -E^B_{\rm v1}$ and  $E^B_{\rm v3} -E^B_{\rm v1}$ 
is relatively small being typically around 0.2 MeV and the magnitude of these two quantities is around 
1 MeV.  Another observation is that the v3 version of separable pairing leads to somewhat higher 
fission barriers as compared with the ones obtained with v2 version. Note that these nuclei are 
extremely neutron-rich with $N/Z$ ratio exceeding 2.0.  Thus, it is reasonable to expect that the 
magnitude of the increase of fission barriers due to isospin dependence of pairing will be lower in 
the nuclei located closer to the $\beta$-stability line.

%%%%%%%%%%%%%%%%%%%%%%%%%%%%%%%%%%%%%%%%%%%%%
\begin{table}
\centering
\caption{Fission barriers heights $E^B$ [in MeV] for the indicated nuclei, functionals and 
versions of separable pairing. The columns 6 and 7 show the increases of fission barrier 
heights for two isospin dependent pairing interactions (v2 and v3) as compared to the one 
obtained  with pairing of constant strength (v1).
}
\label{tab-FB}   
\begin{tabular}{|c|c|c|c|c|c|c|}
\hline
Nucleus        & CEDF       & $E^B_{\rm v1}$ & $E^B_{\rm v2}$  & $E^B_{\rm v3}$ & $E^B_{\rm v2} - E^B_{\rm v1}$ & $E^B_{\rm v3} - E^B_{\rm v1}$ \\  \hline
     1              &  2              &       3              &  4                   &  5                 & 6                                        & 7                                        \\  \hline
                     & NL3$^*$   & 5.71               & 6.49                &  6.38           & 0.78                                   & 0.756                                 \\
$^{316}$Fm  & DD-PC1   & 8.83                & 9.11                &  9.45           & 0.28                                   &  0.62                                  \\
                     & DD-ME2   & 9.77                & 10.36             &  10.53         & 0.59                                   &   0.76                                 \\ \hline
%                     & DD-MEY   &                        & 9.06               &                    &                                           &                                           \\  \hline
                     & NL3$^*$    & 6.39               &  7.06               &  7.23           & 0.67                                   & 0.84                                   \\
$^{324}$Rf    & DD-PC1   & 9.08                & 10.44              & 10.58          & 1.36                                   & 1.50                                   \\
                     & DD-ME2   & 10.44             & 10.69              & 12.01          & 0.25                                    & 1.57                                   \\ \hline
%                    & DD-MEY   &                        &10.09               &                    &                                            &                                           \\  \hline
\end{tabular}
\end{table}
%%%%%%%%%%%%%%%%%%%%%%%%%%%%%%%%%%%%%%%%%%%%%%

%%%%%%%%%%%%%%%%%%%%%%%%%%%%%%%%
\section{Conclusions}
\label{sec-2}
%%%%%%%%%%%%%%%%%%%%%%%%%%%%%%%%

   The detailed analysis of the ground state and fission properties of actinides
and superheavy nuclei  important for the $r$ process modeling has been performed 
within the framework of CDFT for the first time (see Ref.\ \cite{TAA.20}). In particular, 
it allowed to establish the systematic uncertainties in the heights of primary fission
barriers and their sources. In addition,  the present investigation reveals isospin 
dependence  of pairing   as an additional factor affecting fission barriers in fission 
cycling regions. Its 
inclusion leads to a substantial increase of fission barriers in very neutron rich nuclei. 
Available covariant \cite{TA.21} and non-relativistic Skyrme \cite{BLS.12,YMSH.12} DFT 
investigations  strongly point to the existence of isospin dependence of effective pairing 
interaction.  However, its details and accurate form  are still under debate.

  This material is based upon work supported by the U.S. Department of Energy, Office 
of Science, Office of Nuclear Physics under Grant No. DE-SC0013037.

\end{document}